\documentclass[aps,twocolumn,showpacs,prl,longbibliography,amsmath,amssymb,floatfix,superscriptaddress]{revtex4-2}

\usepackage{color}
\usepackage{bbm}
\usepackage{graphicx}
\usepackage{dcolumn}
\usepackage{bm}
\usepackage{array}
\usepackage{float}
\usepackage{supertabular}
\usepackage{longtable}
\usepackage{mathrsfs}
\usepackage{txfonts}
\usepackage[bbgreekl]{mathbbol}
\usepackage{wasysym}
\usepackage[T1]{fontenc}
\usepackage[usenames,dvipsnames]{xcolor}
\usepackage{amsmath,amssymb}
\usepackage[colorlinks=true,citecolor=Cerulean,linkcolor=RubineRed,urlcolor=Cerulean]{hyperref}
\usepackage{braket}
\usepackage{comment}

\usepackage[mode=buildnew]{standalone}

\begin{document}

\title{Classical Cellular Automaton for Measurement-Only Entanglement Transitions}
\author{Will Holdhusen}
\affiliation{Department of Physics and Astronomy, Western Washington University, Bellingham, Washington 98225, USA}
\author{Bailey M. McAmis}\affiliation{Department of Physics and Astronomy, Western Washington University, Bellingham, Washington 98225, USA}
\author{Armin Rahmani}
\affiliation{Department of Physics and Astronomy, Western Washington University, Bellingham, Washington 98225, USA}
\affiliation{Advanced Materials Science and Engineering Center, Western Washington University, Bellingham, Washington 98225, USA}
\affiliation{Kavli Institute for Theoretical Physics, University of California, Santa Barbara, California 93106, USA}

\date{September 23, 2026}

\begin{abstract}
We introduce a generalized classical long-range stochastic cellular automaton that exactly captures the entanglement dynamics and transitions of a measurement-only monitored quantum system. The corresponding quantum model consists of a one-dimensional qubit chain subject to competing single-site and long-range Bell measurements, with separations drawn from a power-law distribution.  At every step of each trajectory, the state remains a tensor product of Bell pairs and unpaired qubits, with bipartite entanglement entropy fully encoded in a classical matching of the sites. This exact representation allows us to solve the dynamics analytically in several limits and to determine the steady-state scaling of the entanglement entropy. We find volume-law, fractal, and area-law regimes. Numerical results show that these regimes persist beyond the analytically solvable limit. Because Bell measurement outcomes change only the Bell-state labels and leave the matching unchanged, all subsystem entropies are outcome independent, avoiding the exponential cost of trajectory postselection often encountered in experimental studies of measurement-induced entanglement dynamics.

\end{abstract}

\maketitle

\textit{Introduction.—}
In conventional monitored quantum circuits, unitary evolution generates and spreads entanglement, while local measurements tend to suppress it, leading to entanglement transitions \cite{LiChenFisher2018,SkinnerRuhmanNahum2019,
LiChenFisher2019,BenZionMcGreevyGrover2020,BaoEtAl2020,ChoiEtAl2020,GullansHuse2020Dynamical}. Measurements of suitable multiqubit observables, however, can also generate entanglement without unitary dynamics \cite{LangBuchler2020,NahumSkinner2020,IppolitiEtAl2021,KlockeBuchhold2022,KlockeBuchhold2023,SriramEtAl2023, Huang2026}, including with long-range measurements \cite{Piccitto2023StringMeasurements, Zhu2026,
Russomanno2023LongRangeMonitoring,Kuno2023,Gomez2026}. These theoretical advances have been accompanied by experimental progress \cite{NoelEtAl2022,HokeEtAl2023}. Several models of nonunitary dynamics, e.g., based on spacetime transformations of quantum circuits, can give rise to states with fractal sub-volume-law entanglement entropy \cite{Ippoliti2022}. Similar scaling has been found in monitored systems with long-range interactions or two-qubit gates \cite{Minato2022,Block2022,SharmaEtAl2022,Muller2022LongRange}
and long-range measurements
~\cite{Russomanno2023LongRangeMonitoring,Kuno2023,Gomez2026}. 

Here we introduce a measurement-only model based on stochastic long-range Bell measurements, joint two-qubit measurements central to quantum teleportation \cite{ZukowskiEtAl1993,BennettEtAl1993}. The model’s entanglement dynamics can be mapped exactly onto a classical stochastic cellular automaton, generalized to include nonlocal updates~\cite{Cannas1998,PizziEtAl2021}. 
Our analytical and numerical results reveal a steady-state phase diagram comprising volume-law, fractal, and area-law entanglement regimes. Furthermore, these calculations provide the time evolution of the entanglement entropy as the system approaches its steady state. While sharing important features with the systems of Refs.~\cite{Block2022,SharmaEtAl2022,NahumSkinner2020,KlockeBuchhold2023}, the present model combines competing single-site and long-range measurements into a closed stochastic dynamics of classical matchings, which allows analytical derivation of entanglement scaling in controlled limits.

Our system consists of a one-dimensional chain of qubits initialized in a product state. At each step, either a single site is measured with probability $p$ or a pair of sites undergoes a Bell measurement with probability $1-p$.
The pair is chosen with weight proportional to $r^{-\alpha}$, where $r$ is the separation between the two qubits. Every trajectory remains a tensor product of Bell pairs and unpaired qubits, so the entanglement structure is fully encoded by a classical matching of the sites. The evolution of the matching defines a long-range generalized cellular automaton with pairing moves, reminiscent of stochastic processes on loop and connectivity configurations \cite{PearceEtAl2002,DeGierNienhuis2005,Nahum2013}.

While the states contain only pairwise rather than generic multipartite entanglement, the model sheds light on the geometric mechanisms by which measurements generate and spread entanglement~\cite{Chan2019UnitaryProjective}. 
Although the dynamics maps onto a stabilizer circuit that may be solved in polynomial time \cite{Aaronson2004, Gidney2021}, the classical automaton is considerably more efficient for numerical simulations since the entanglement structure is encoded entirely by the pairing configuration, enabling each update to be performed in $\mathcal O(1)$ time as in Ref.~\cite{KlockeBuchhold2023}. 

Our measurement protocol may be amenable to experimental realizations. Bell measurements may be realized on quantum hardware through ancilla-assisted measurements of stabilizers in superconducting circuits \cite{AndersenEtAl2019} or by combining entangling gates with local readout, as in trapped-ion teleportation experiments \cite{RiebeEtAl2004}.
The model also avoids the postselection problem, which complicates experimental probes of monitored entanglement: nonlinear trajectory observables require repeated realizations of the same, exponentially unlikely measurement record \cite{GullansHuse2020Scalable}. Proposed workarounds rely on decoding, cross-entropy benchmarks, feedback, classical shadows, or special circuit structures \cite{BuchholdMullerDiehl2022,LiEtAl2023CrossEntropy,HokeEtAl2023,GarrattAltman2024Postselection,PassarelliEtAl2024}. In our model, no postselection on measurement outcomes is required: Bell outcomes affect only the Bell-state labels, while the pairing configuration, and hence every subsystem entropy, is determined solely by which sites are measured. 

\begin{figure}
\includegraphics*[width=.9\columnwidth]{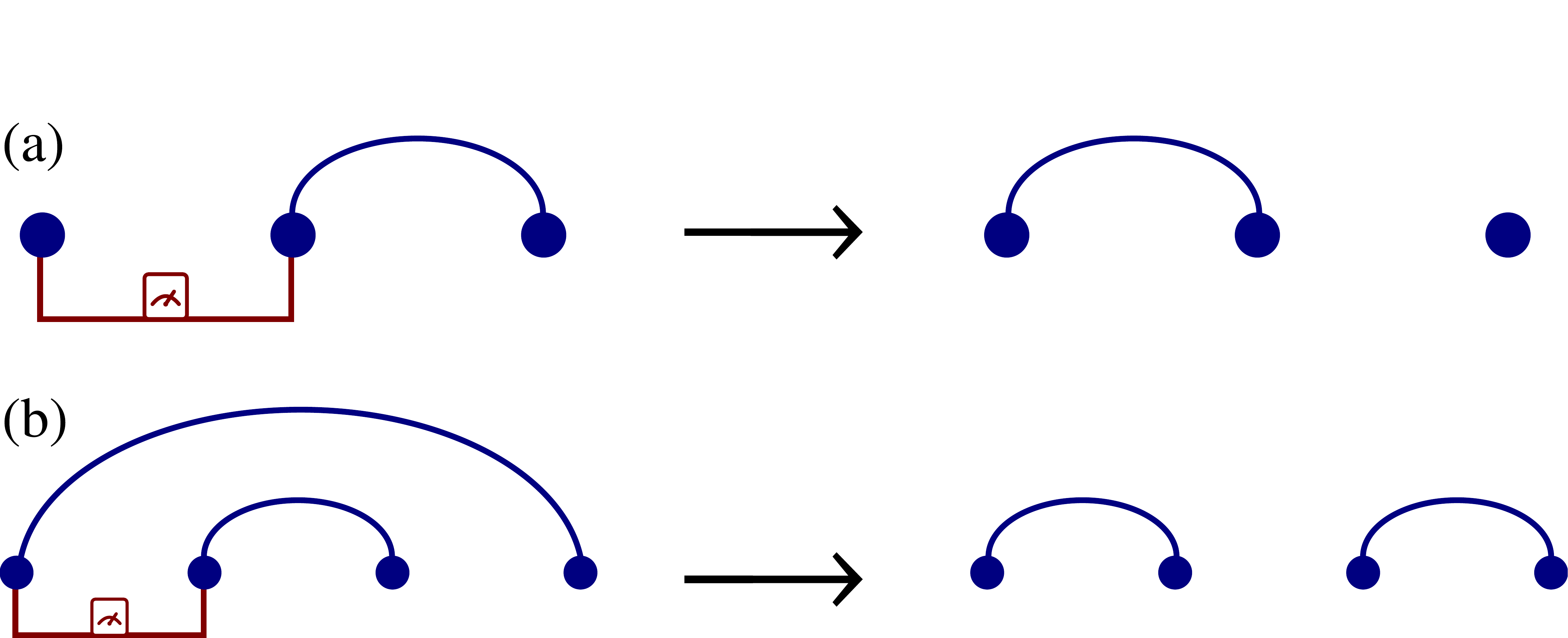}
\caption{Elementary moves of the stochastic cellular automaton. Red links and meter symbols indicate Bell measurements, while blue arcs denote Bell pairs. (a) Measuring an unpaired qubit and one member of a Bell pair replaces the original pair and leaves its former partner unpaired. (b) Measuring one qubit from each of two Bell pairs produces entanglement swapping, reconnecting their former partners.
\label{fig:1}}
\end{figure}

\textit{Model.--- }
We consider $N$ qubits initially prepared in the product state $|\psi_0\rangle=|00\cdots0\rangle$. For simplicity, we arrange the qubits on a one-dimensional lattice with open boundary conditions, although the construction extends naturally to higher dimensions, other lattice geometries, and irregular spatial arrangements.
We then apply Bell measurements to 
randomly selected pairs of qubits chosen such that the probability of measuring sites $i$ and $j$ is proportional to $|i-j|^{-\alpha}$.
Each Bell measurement projects the selected pair onto one of the four orthonormal maximally-entangled Bell states, $|\psi_\pm\rangle=\left(|01\rangle\pm|10\rangle\right)/\sqrt{2}$ and $|\phi_\pm\rangle=\left(|00\rangle\pm|11\rangle\right)/\sqrt{2}$.

In addition to the entangling Bell measurements, the model includes single-qubit measurements in the computational basis, which project the selected qubit onto $|0\rangle$ or $|1\rangle$. If the qubit belongs to a Bell pair, the measurement breaks the pair, projecting its partner onto a definite computational-basis state and leaving both qubits unpaired.

As the dynamics proceeds, a Bell measurement may act on qubits that already belong to Bell pairs. First, suppose that qubit $a$ is unpaired and in the state $|0\rangle$ or $|1\rangle$, while qubits $b$ and $c$ form a Bell pair. As an example for the unpaired qubit $a$ in state $|0\rangle$ and $b$ and $c$ in $|\psi_\pm\rangle$ , expressing the initial state in the Bell basis gives
\begin{equation}
\begin{split}
|0\rangle_a|\psi_\pm\rangle_{bc}
=\frac{1}{2}\big[&
\left(|\phi_+\rangle_{ab}+|\phi_-\rangle_{ab}\right)|1\rangle_c\\
&\pm
\left(|\psi_+\rangle_{ab}+|\psi_-\rangle_{ab}\right)|0\rangle_c
\big].\label{eq:bell-single}
\end{split}
\end{equation}
A Bell measurement on $a$ and $b$ therefore pairs these two qubits and projects $c$ onto $|0\rangle$ or $|1\rangle$. At the level of the classical matching, the bond $(b,c)$ is replaced by $(a,b)$, leaving $c$ unpaired. The other possible initial states have a similar structure as shown in Sec. I of the Supplemental Material (SM) \cite{SM}.

If both measured qubits already belong to Bell pairs, the measurement produces entanglement swapping. Suppose that $a$ is paired with $c$ and $b$ with $d$. For example,
\begin{equation}
\begin{split}
|\phi_+\rangle_{ac}|\phi_+\rangle_{bd}
=\frac{1}{2}\big(&
|\phi_+\rangle_{ab}|\phi_+\rangle_{cd}
+|\phi_-\rangle_{ab}|\phi_-\rangle_{cd}\\
&+|\psi_+\rangle_{ab}|\psi_+\rangle_{cd}
+|\psi_-\rangle_{ab}|\psi_-\rangle_{cd}
\big).
\end{split}
\label{eq:bell-bell}
\end{equation}
This identity explicitly shows that a Bell measurement on $(a,b)$ projects $(c,d)$ onto a Bell state. All 16 possible products of initial Bell states admit an analogous decomposition (see SM). For each of the four measurement outcomes on $(a,b)$, the remaining qubits $(c,d)$ are projected onto a definite Bell state. Thus, although the Bell-state label of $(c,d)$ depends on the initial states and the measurement outcome, the pairing update is always $(a,c),(b,d)\longrightarrow(a,b),(c,d)$.
The evolution of the matching is therefore independent of the Bell-measurement outcome. These moves are shown in Fig.~\ref{fig:1}.

Using natural logarithms, each Bell pair crossing the boundary of a subsystem contributes \(\ln 2\) to its entanglement entropy. Alternatively, by defining the entropy as
$S_A(N_A)=-{\rm Tr}\!\left(\rho_A\log_2\rho_A\right)
$,
each Bell pair connecting $A$ to its complement contributes exactly one unit. Thus, $S_A(N_A)$ is simply the number of Bell pairs connecting $A$ to its complement. We adopt this convention throughout.

Since entanglement only depends on the matching of Bell pairs, the state's complete entanglement structure may be encoded as a vector of $N$ integers where the $j$th entry either marks the site as unentangled or identifies its Bell partner. Each single-qubit or Bell measurement updates this vector in constant time.
Encoding the full quantum state would require further variables specifying the single-qubit and Bell states, with update rules derived from decompositions such as Eqs.~\eqref{eq:bell-single} and \eqref{eq:bell-bell}.

The automaton allows for direct simulation of the configuration vector described above. At each step, a single-qubit measurement occurs with probability $p$, where a randomly chosen site and its Bell partner, if any, become unentangled. Otherwise, a Bell measurement acts on a pair selected by sampling its separation $r$ from an input pair-length distribution $P_\text{in}(r)$ using a Walker–Vose alias table \cite{Walker1974,Vose1991}, then choosing among pairs at that separation. For open boundary conditions, $P_\text{in}(r)\propto (N-r)r^{-\alpha}$ accounts for the $N-r$ pairs of length $r$, thus ensuring the weight for each pair goes as $r^{-\alpha}$. In the numerical simulation, we generate many measurement sequences. Unless otherwise stated, the entanglement entropy $S_A$, unpaired qubit density $n_1$, and steady-state pair length distribution $P(r)$ refer to averages over measurement trajectories.
In our simulations, we achieved adequate convergence with $10^3$ to $10^6$ realizations. 

\begin{figure}
\includegraphics[width=\linewidth]{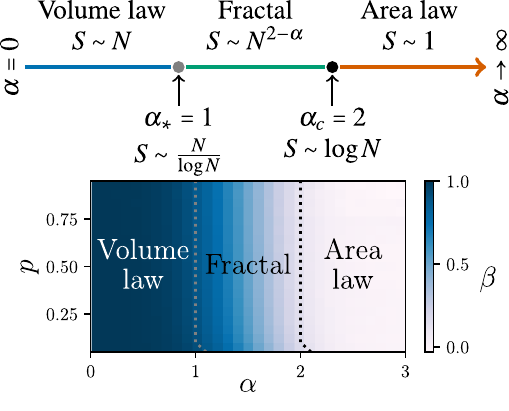}
\caption{
Top: analytically derived entanglement phase diagram in the dilute-pair ($p\rightarrow 1^-$) limit. Bottom: phase diagram for $p>0$ obtained from simulations. Heatmap color indicates entropy scaling exponent $\beta$ where $S\sim N^\beta$. The limit of $p=0$ is challenging to study numerically, but analytical results in the $\alpha\to\infty$
limit give logarithmic entanglement scaling. At $p=1$, the model is fully unentangled and is therefore trivially area law.}
\label{fig:full_phase_diagram}
\end{figure}

\textit{Entanglement phase diagram.—}
The steady-state entanglement is analytically tractable in two limits: the all-to-all limit $\alpha=0$, in which every pair is selected with equal probability, and the dilute-pair limit $p\to1^{-}$, in which pair--pair encounters are rare and the steady-state bond-length distribution
follows the sampling distribution $P_\text{in}(r)$. Combining these results with simulations at intermediate $p$ yields the phase diagram in Fig.~\ref{fig:full_phase_diagram}. We analyze the two solvable limits before presenting the numerical results.

At $\alpha=0$, pair selection is distance independent, so every pair of sites is sampled with equal probability. This leads to a spatially-uniform distribution of paired and unpaired states. Under these conditions, it can be seen that at each step in the simulation, the number of unentangled sites $N_1$ can either increase by two with probability $p\left(1-\frac{N_1}{N}\right)$ in the case that a single-qubit measurement is applied to an entangled qubit, remain the same, or decrease by two with probability $(1-p)\frac{N_1}{N}\frac{N_1-1}{N-1}$. Neglecting terms of $\mathcal O(1/N)$ and approximating time as continuous leads to the differential equation
\begin{equation}
\frac{dn_1}{dt}
=\frac{2}{N}\left[p(1-n_1)-(1-p)n_1^2\right],
\quad n_1=\frac{N_1}{N}.
\label{eqn:dndt}
\end{equation} 
For $0<p<1$ and $n_1(0)=1$, the solution is
\begin{equation}
n_1(t)=n_-+\frac{n_+-n_-}{1-\frac{1-n_+}{1-n_-}e^{-t/\tau}},
\qquad
n_\pm=\frac{-p\pm\sqrt{4p-3p^2}}{2(1-p)},
\label{eqn:continuum_single}
\end{equation}
with
the relaxation time 
\begin{equation}
    \tau=\frac{N}{2\sqrt{4p-3p^2}}. \label{eqn:tau}
\end{equation} 
The steady-state value is given by $\lim_{t\to\infty}n_1(t)=n_+$. At $p=0$, Eq.~\eqref{eqn:dndt} reduces to $\frac{dn_1}{dt}=-\frac{2}{N}n_1^2$, giving the algebraic decay $n_1(t)=1/(1+2t/N)$.
At the opposite endpoint $p=1$, no Bell pairs are generated and $n_1(t)=1$.

In addition to $n_1$, the subsystem entanglement entropy $S_A(N_A)$ can also be computed directly when $\alpha=0$ by accounting for the probabilities of measurement processes that change it.
A single-site measurement on either endpoint of a crossing pair destroys that pair, giving $\Delta S_A(N_A)=-1$ with probability $2p(1-n_1)n_A(1-n_A)$ where $n_A=N_A/N$. Conversely, a Bell measurement on two unpaired sites on opposite sides of the bipartition creates a crossing pair, giving $\Delta S_A(N_A)=+1$ with probability $2(1-p)n_1^2n_A(1-n_A)$. 
Other Bell-measurement processes can also change the entropy, but for uniform pair selection at $\alpha=0$, their contributions cancel on average.
Therefore, we obtain
\begin{equation}
    \frac{dS_A(N_A)}{dt}
=2n_A(1-n_A)
\left[(1-p)n_1^2-p(1-n_1)\right].
\end{equation}
Comparison with Eq.~\eqref{eqn:dndt} gives $\frac{dS_A(N_A)}{dt}
=-N n_A(1-n_A)\frac{dn_1}{dt}$, which integrates to
\begin{equation}\label{eqn:continuum_entropy}
S_A(N_A,t)=N n_A(1-n_A)(1-n_1(t))
\end{equation}
using the initial condition $n_1(0)=1$.
Thus, apart from the trivial case $p=1$, the model at $\alpha=0$ exhibits volume-law entanglement scaling. For $p>0$, the entanglement  entropy converges to its steady state with relaxation time $\tau$. The steady-state behavior for $\alpha=0$ can also be studied using a combinatorial approach discussed in Sec. II of the SM~\cite{SM}.

\begin{figure}
\includegraphics[width=\linewidth]{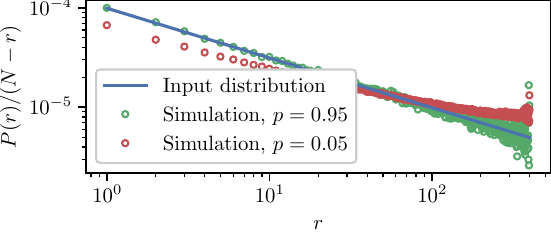}
\caption{Measured and input pair distributions at $\alpha=0.5$ for a 400-qubit system averaged over $10^5$ trials.
These results indicate good agreement between input and measured distributions in the dilute limit ($p=0.95$). When $p$ is reduced to $0.05$, entanglement swapping increases the tail of the distribution.}
\label{fig:pair_distributions}
\end{figure}

We now consider the dilute-pair limit, $p=1-\epsilon$ with $\epsilon\ll 1$. Because unentangling measurements occur with high probability, the steady state is expected to have a low pair density, $\rho=\frac{1}{2}(1-n_1)$. Consequently, the probability of performing an entangling measurement on two qubits that are already entangled is small, $\mathcal O\left(\epsilon^2\right)$. Since such events are precisely what generate pairs whose lengths are not directly drawn from the input distribution $P_\text{in}(r)$, the steady-state pair-length distribution $P(r)$ should approach $P_\text{in}(r)$ in this limit. Simulation results shown in Fig.~\ref{fig:pair_distributions} confirm this expectation. At $p=0.95$, the measured output distribution closely follows the input distribution, while at $p=0.05$, the measured distribution shows a distinct enhancement of long-range pairs due to entanglement swapping.

\begin{figure}
\includegraphics[width=\linewidth]{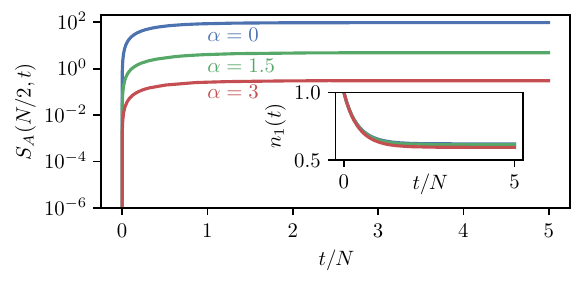}
\caption{Half-chain entropy as a function of time in a system with $N=1000$ and unentangling measurement rate $p=0.5$ averaged over $10^5$ trials.
Inset: density of unpaired qubits as a function of time for the same parameters.
}
\label{fig:dynamics}
\end{figure}

Using $P(r)\approx P_\text{in}(r)$ in the dilute-pair limit, we can derive the average entanglement entropy $S_A(N_A)$ for an open chain; periodic boundary conditions are considered in Sec. III of the SM \cite{SM}. 
A pair of sites separated by a distance $r$ has a weight proportional to $r^{-\alpha}$. Since there are $N-r$ different pairs of length $r$, the probability of any particular pair of length $r$ is $r^{-\alpha}/Z_N$, where $Z_N=\sum_{r=1}^{N-1}(N-r)r^{-\alpha}$. 
If the subsystem $A$ consists of the left $N_A$ sites of the chain, the number of bonds of length $r$ connecting $A$ to its complement is $C(r,N_A)=\min\left(r,N_A,N-N_A,N-r\right)$ (it must be bounded by $N_A$, $N-N_A$, $N-r$, and $r$ as a bond of length $r$ crosses the boundary only if its left endpoint lies within $r$ sites of the boundary). 
If $N_1=n_1N$ sites are unpaired, the total number of bonds is $(N-N_1)/2=N(1-n_1)/2$.
The average number of bonds connecting $A$ to its complement, and hence the entanglement entropy, is therefore
\begin{equation}
S_A(N_A)=\frac{N(1-n_1)}{2Z_N}
\sum_{r=1}^{N-1}
\min\left(r,N_A,N-N_A,N-r\right)r^{-\alpha}.
\label{eqn:SNA}
\end{equation}

The leading large-$N$ behavior follows from an Euler--Maclaurin expansion of the sum. We find that in the dilute-pair limit and for $\alpha\neq 1,2$, the half-chain entropy $S_A(N/2)$ scales as
\begin{equation}
S_A(N/2)\sim N^\beta
\label{eqn:entropy_power_law}.
\end{equation}
For $\alpha<1$, $\beta=1$ giving volume-law entanglement scaling. When $1<\alpha<2$, we obtain fractal (subvolume-law) scaling with $\beta=2-\alpha$. When $\alpha>2$, the steady state has area-law (constant) entanglement scaling with $\beta=0$.
These entanglement phases are separated by transitions at $\alpha_*=1$ and $\alpha_c=2$, with scaling $S\sim N/\log N$ at $\alpha_*$ and $S\sim \log N$ at $\alpha_c$.
The corresponding prefactors and subleading corrections are presented in Sec. III of the SM, along with analysis of the case of periodic boundary conditions showing the same leading-order scaling \cite{SM}. The analytical scaling of the entanglement entropy in the dilute-pair limit is shown in the top panel of Fig.~\ref{fig:full_phase_diagram}. 

\textit{Numerical Simulations.—}
Connecting these analytical results and extending the entanglement phase diagram to finite $p$ and $\alpha$ requires numerical simulations. Eqs.~\eqref{eqn:continuum_single} and \eqref{eqn:continuum_entropy} suggest rapid convergence to a steady state in the case $\alpha=0$. Numerical results, with an example shown in Fig.~\ref{fig:dynamics}, confirm this and further indicate that the relaxation time for a given system size depends only weakly on $\alpha$ for $p>0$.
To ensure well-converged steady states, our $p>0$ simulations were evolved for $T=200N$ timesteps, well beyond the relaxation time $\tau$ given by Eq.~\eqref{eqn:tau}. 
Even for $p\sim 10^{-2}$, $\tau/N\sim \mathcal O(1)$.

\begin{figure}
\includegraphics[width=\linewidth]{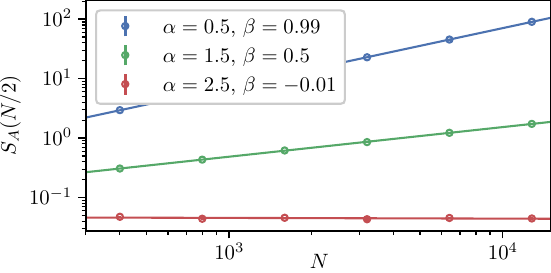}
\caption{Log-log plots of $S_A(N/2)$ obtained from simulations with $p=0.95$ averaged over $10^5$ trials. Dashed lines correspond to least-squares fits to a power law $A N^\beta$. Error bars (not visible for most points due to their small size) correspond to the standard error of the mean (SEM) from these trials.}
\label{fig:exponent_examples}
\end{figure}

To compare our simulation results with the entropy scaling shown in Fig.~\ref{fig:full_phase_diagram}, we fit $S_A(N/2)$ to the form Eq.~\eqref{eqn:entropy_power_law} and extract entropy-scaling exponent $\beta$,
finding exponents consistent with the dilute-limit predictions at $p=0.95$ as depicted in Fig.~\ref{fig:exponent_examples}. Simulations at varying $p$ show that the exponents predicted in the dilute limit persist with only slight changes even for very small finite $p$.
These results are shown in the heatmap in Fig.~\ref{fig:full_phase_diagram}, where exponents were extracted from finite-size fits of system sizes ranging from 400 to 12800 qubits, averaged over $10^4$ trials. A more detailed exploration of these exponents in the thermodynamic limit is presented in Sec. IV of the SM \cite{SM}.

\textit{Detection of entanglement transitions.—}
The volume-law, fractal, and area-law phases of our model are separated by entanglement transitions at $\alpha_*=1$ and $\alpha_c=2$ in the dilute-pair limit.
In practice, significant finite-size effects, discussed in Sec. V of the SM \cite{SM}, obscure the expected ($S\sim N^\beta$) entanglement scaling near these transitions, making their precise location difficult to determine from the scaling exponent. We therefore identify the transitions by searching for the characteristic entanglement scaling expected at $\alpha_*$ and $\alpha_c$. Specifically, the Euler--Maclaurin expansion of the dilute-pair entropy, Eq.~\eqref{eqn:SNA}, to first subleading order at $\alpha=1$ suggests a finite-size form 
$
S_A(N/2) = A\frac{N}{\log N} + B \frac{N}{\left(\log N\right)^2}
$ for $\alpha_*$. Likewise, at $\alpha=2$ the expansion suggests
$
S_A(N/2) = A\log N + B
$
for $\alpha_c$.
For each $p$, we perform fits to the expected scaling forms on a grid of $\alpha$ values with spacing $\Delta \alpha = 0.01$, identifying the transition with the value of $\alpha$ that maximizes the coefficient of determination $R^2$. The resulting transition locations are therefore reported up to the grid spacing. 

The phase boundaries depart appreciably from their dilute-limit values only for $p\lesssim 0.1$. At $\alpha=0$, Eq.~\eqref{eqn:continuum_entropy} predicts volume-law entanglement for every $0\leq p<1$, including $p=0$. In the nearest-neighbor limit $\alpha\to\infty$, the $p=0$ dynamics are related to Majorana loop models with logarithmic entanglement scaling \cite{NahumSkinner2020,KlockeBuchhold2023}. Resolving the behavior at finite $\alpha$ and $p=0$ numerically is difficult because relaxation is slow: $n_1(t)$ decays algebraically at $\alpha=0$, as discussed after Eq.~\eqref{eqn:continuum_single}, and slow dynamics also occur in the nearest-neighbor limit \cite{NahumSkinner2020}. Consequently, apparent area-law behavior in finite-time simulations may just be an artifact. The results presented in Sec. VI of the SM suggest that such behavior at $p=0$ could give way to logarithmic scaling or fractal scaling with a small exponent at longer times \cite{SM}. However, beyond the $\alpha=0$ and $\alpha\to\infty$ limits, the asymptotic entanglement scaling on this line remains unresolved for the times and system sizes studied.

\textit{Conclusions.—}
We have shown that the entanglement dynamics of a measurement-only qubit chain maps exactly onto a classical stochastic matching process. Pairs of qubits subjected to Bell measurement are selected according to a power-law distribution $r^{-\alpha}$. In the dilute-pair limit, the half-chain entropy follows a volume law for $\alpha<1$, a fractal power law for $1<\alpha<2$, and an area law for $\alpha>2$, with logarithmic corrections at phase boundaries. Numerically, all three regimes persist well beyond this analytically tractable limit. The classical cellular automaton enables analytical studies and highly efficient numerical simulations of entanglement transitions. Future work could extend the model to higher dimensions and explore other pair-selection distributions and spatially disordered measurement rates. Because the matching depends on the measured sites but not on Bell-measurement outcomes, subsystem entropies require no postselection on those outcomes. This offers a practical route to experimentally probing measurement-induced entanglement dynamics and transitions.

\section{Acknowledgements}
We thank Sidney Redner for helpful discussions. This work was primarily supported by the U.S. Department of Energy under Grant No. DE-SC0024641. A.R. acknowledges the hospitality of the Kavli Institute for Theoretical Physics, supported by NSF Grant No. PHY-2309135, and the Aspen Center for Physics, supported by NSF Grant No. PHY-1607611.
\bibliography{references}

\end{document}


\title{Classical Cellular Automaton for Measurement-Only Entanglement Transitions: Supplemental Material}
\author{Will Holdhusen}
\affiliation{Department of Physics and Astronomy, Western Washington University, Bellingham, Washington 98225, USA}
\author{Bailey M. McAmis}\affiliation{Department of Physics and Astronomy, Western Washington University, Bellingham, Washington 98225, USA}
\author{Armin Rahmani}
\affiliation{Department of Physics and Astronomy, Western Washington University, Bellingham, Washington 98225, USA}
\affiliation{Advanced Materials Science and Engineering Center, Western Washington University, Bellingham, Washington 98225, USA}
\affiliation{Kavli Institute for Theoretical Physics, University of California, Santa Barbara, California 93106, USA}

\date{September 23, 2026}

\maketitle

\section{Bell-pair labels in the automaton moves}
Here, we explore all possible outcomes of a Bell measurement on sites $a$ and $b$ under the assumption that the full quantum state is a tensor product of Bell pairs and unentangled qubits projected into the computational basis.

Suppose that $a$ is unentangled in the state $|0\rangle$ or $|1\rangle$, while $b$ forms a Bell pair with $c$. Expanding the three-qubit state in the Bell basis of $(a,b)$ gives
\begin{eqnarray*}
|0\rangle_a|\psi_\pm\rangle_{bc}
&=&\frac{1}{2}\left[
\left(|\phi_+\rangle_{ab}+|\phi_-\rangle_{ab}\right)|1\rangle_c
\pm
\left(|\psi_+\rangle_{ab}+|\psi_-\rangle_{ab}\right)|0\rangle_c
\right],\\
|1\rangle_a|\psi_\pm\rangle_{bc}
&=&\frac{1}{2}\left[
\left(|\psi_+\rangle_{ab}-|\psi_-\rangle_{ab}\right)|1\rangle_c
\pm
\left(|\phi_+\rangle_{ab}-|\phi_-\rangle_{ab}\right)|0\rangle_c
\right],\\
|0\rangle_a|\phi_\pm\rangle_{bc}
&=&\frac{1}{2}\left[
\left(|\phi_+\rangle_{ab}+|\phi_-\rangle_{ab}\right)|0\rangle_c
\pm
\left(|\psi_+\rangle_{ab}+|\psi_-\rangle_{ab}\right)|1\rangle_c
\right],\\
|1\rangle_a|\phi_\pm\rangle_{bc}
&=&\frac{1}{2}\left[
\left(|\psi_+\rangle_{ab}-|\psi_-\rangle_{ab}\right)|0\rangle_c
\pm
\left(|\phi_+\rangle_{ab}-|\phi_-\rangle_{ab}\right)|1\rangle_c
\right].
\end{eqnarray*}
It is clear from the general structure above that a Bell measurement on $a$ and $b$ projects $c$ onto a single-qubit state in the computational basis.

Now suppose that $a$ is entangled with $c$ and $b$ with $d$. In the example given in the main text, we considered $|\phi_+\rangle_{ac}|\phi_+\rangle_{bd}
=\frac{1}{2}\big(
|\phi_+\rangle_{ab}|\phi_+\rangle_{cd}
+|\phi_-\rangle_{ab}|\phi_-\rangle_{cd}
+|\psi_+\rangle_{ab}|\psi_+\rangle_{cd}
+|\psi_-\rangle_{ab}|\psi_-\rangle_{cd}
\big)$. The decomposition of the 16 possible products of Bell states on $(a,c)$ and $(b,d)$ in the Bell basis of $(a,b)$ and $(c,d)$ is summarized in Table \ref{tab:bell-swapping}.
The rows are the products of Bell states for $(a,c)$ and $(b,d)$. The columns specify the
Bell states of $(a,b)$, and the entries 
represent the term multiplying it in the $(c,d)$ Bell basis, with a common
coefficient $1/2$. The first row corresponds to the example of the main text. We see from the above decomposition that each Bell-measurement on $(a,b)$ projects $(c,d)$ onto a definite Bell state, leading to the dynamical moves discussed in the main text.

\begin{table}[h]
\centering
\small
\renewcommand{\arraystretch}{1.15}
\begin{tabular*}{\columnwidth}{
@{\extracolsep{\fill}}c|cccc@{}
}
\hline\hline
$(ac, bd)$
& $\phi_+$
& $\phi_-$
& $\psi_+$
& $\psi_-$\\
\hline
$\phi_+\phi_+$ & $+\phi_+$ & $+\phi_-$ & $+\psi_+$ & $+\psi_-$\\
$\phi_+\phi_-$ & $+\phi_-$ & $+\phi_+$ & $-\psi_-$ & $-\psi_+$\\
$\phi_+\psi_+$ & $+\psi_+$ & $+\psi_-$ & $+\phi_+$ & $+\phi_-$\\
$\phi_+\psi_-$ & $+\psi_-$ & $+\psi_+$ & $-\phi_-$ & $-\phi_+$\\
\hline
$\phi_-\phi_+$ & $+\phi_-$ & $+\phi_+$ & $+\psi_-$ & $+\psi_+$\\
$\phi_-\phi_-$ & $+\phi_+$ & $+\phi_-$ & $-\psi_+$ & $-\psi_-$\\
$\phi_-\psi_+$ & $+\psi_-$ & $+\psi_+$ & $+\phi_-$ & $+\phi_+$\\
$\phi_-\psi_-$ & $+\psi_+$ & $+\psi_-$ & $-\phi_+$ & $-\phi_-$\\
\hline
$\psi_+\phi_+$ & $+\psi_+$ & $-\psi_-$ & $+\phi_+$ & $-\phi_-$\\
$\psi_+\phi_-$ & $-\psi_-$ & $+\psi_+$ & $+\phi_-$ & $-\phi_+$\\
$\psi_+\psi_+$ & $+\phi_+$ & $-\phi_-$ & $+\psi_+$ & $-\psi_-$\\
$\psi_+\psi_-$ & $-\phi_-$ & $+\phi_+$ & $+\psi_-$ & $-\psi_+$\\
\hline
$\psi_-\phi_+$ & $-\psi_-$ & $+\psi_+$ & $-\phi_-$ & $+\phi_+$\\
$\psi_-\phi_-$ & $+\psi_+$ & $-\psi_-$ & $-\phi_+$ & $+\phi_-$\\
$\psi_-\psi_+$ & $-\phi_-$ & $+\phi_+$ & $-\psi_-$ & $+\psi_+$\\
$\psi_-\psi_-$ & $+\phi_+$ & $-\phi_-$ & $-\psi_+$ & $+\psi_-$\\
\hline\hline
\end{tabular*}
\caption{Decomposition of the product of Bell states pairing qubits $(a,c)$ and $(b,d)$ in the Bell basis of $(a,b)$ and $(c,d)$. }
\label{tab:bell-swapping}
\end{table}

The general update above can be expressed compactly using binary labels. Define $|\beta_{xz}\rangle=(I\otimes X^xZ^z)|\phi_+\rangle$, where $(x,z)=(0,0),(0,1),(1,0),(1,1)$ label $|\phi_+\rangle$, $|\phi_-\rangle$, $|\psi_+\rangle$, and $|\psi_-\rangle$, respectively. Suppose that $(a,c)$ and $(b,d)$ are initially in the states $|\beta_{x_1z_1}\rangle$ and $|\beta_{x_2z_2}\rangle$. If a Bell measurement on $(a,b)$ yields $|\beta_{x_mz_m}\rangle$, the remaining qubits are projected, up to an irrelevant overall phase, onto
$|\beta_{x_1\oplus x_2\oplus x_m,
z_1\oplus z_2\oplus z_m}\rangle_{cd}$.
Depending on the state of the qubits before the measurement, the Bell measurement on $a$ and $b$ leads to the following update of the pairing structure  \begin{align*}
&a,b \rightarrow (a,b) \\
&a,(b,c) \rightarrow (a,b),c\\
&(a,c),(b,d)\rightarrow (a,b),(c,d).
\end{align*}

\section{$\alpha=0$ Combinatorial analysis}
In the main text, we derive the $\alpha=0$ entanglement entropy directly from the update rules. Here, we present an alternative combinatorial derivation of the steady-state entropy by averaging over uniformly distributed pairing configurations and assuming independence of unoccupied sites, thus providing an independent check and making explicit the underlying pairing statistics.
\begin{figure}
    \centering
    \includegraphics[width=0.95\linewidth]{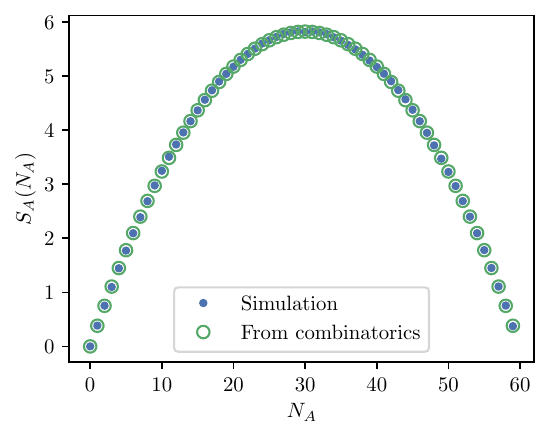}
    \caption{Entropy (bonds across cuts) calculated in a 60 qubit system with single-site measurement probability $p=0.5$ averaged over $10^4$ seeds after 1000 timesteps.
    Open circles indicate results obtained from Eq.~\eqref{eqn:combinatoric_entropy}.}
    \label{fig:combinatoric_entropy}
\end{figure}

Consider a bipartition of the model into subsystems $A$ and $B$ consisting of $N_A$ ($N_B$) sites. Let $x_A$ ($x_B$) count the number of entangled sites in these subsystems, with $M=\frac{1}{2}(x_A+x_B)$ bonds connecting the entangled sites.
Let $L$ be the number of bonds connecting $A$ and $B$ ($L=S_A(N_A)$). This leaves $x_A-L$ sites in $A$ with their partners also in $A$, and similarly for $B$. Note this means $x_A-L$ and $x_B-L$ must both be even numbers. Writing $2y_A = x_A-L$ and $2y_B= x_B-L$, it can be seen that there are 
\begin{equation*}
F(2y)=\binom{2y}{y}\frac{y!}{2^y}=(2y-1)!!
\end{equation*}
distinct ways of pairing $2y$ sites. So, for fixed $x_A$ and $x_B$, there are
\begin{align*}
G(x_A,x_B, L)=&\binom{x_A}{L}\binom{x_B}{L}F(x_A-L)F(x_B-L)L!\\
=&{x_A!x_B!\over L! (x_A-L)!!(x_B-L)!!}
\end{align*}
configurations with $L$ bonds connecting $A$ with $B$.
Then, for fixed $x_A$ and $x_B$, the probability of having $L$ entangling bonds is
\[
P(x_A,x_B,L)={G(x_A,x_B, L)\over \sum_{{\rm allowed}\: L'}G(x_A,x_B, L')}.
\]
For a bipartition of a 1-d chain,
\[
{\rm allowed }\:L=\begin{cases}
0, 2, ...\min(x_A,x_B), &  x_{A,B}\:{\rm even}, \\ 
1, 3, ...\min(x_A,x_B), &  x_{A,B}\:{\rm odd}.
\end{cases}
\]
The number of configurations with $x_A$ and $x_B$ paired sites in the subsystems is
\[
N(x_A,x_B)=\binom{N_A}{x_A}\binom{N_B}{x_B}.
\]
and the probability of each is determined by the probability of a site belonging to a pair, $(1-n_1)$, where $n_1$ is the unpaired-qubit density given by Eq.~(3) in the main text.
Since the paired sites are uniformly distributed, the $N(x_A,x_B)$ configurations occur with probability $(1-n_1)^{x_A+x_B} n_1^{N_A+N_B-x_A-x_B}$.

Putting all of this together allows us to calculate average entanglement entropy $S_A(N_A)=\braket L$:
\begin{widetext}
\begin{equation}
S_A(N_A)={\sum'_{x_A,x_B} \sum_{{\rm allowed}\: L} 
L P(x_A,x_B,L)N(x_A,x_B)(1-n_1)^{x_A+x_B}n_1^{N_A+N_B-x_A-x_B}\over \sum'_{x_A,x_B} \sum_{{\rm allowed}\: L}  P(x_A,x_B,L)N(x_A,x_B)(1-n_1)^{x_A+x_B}n_1^{N_A+N_B-x_A-x_B}}
\label{eqn:combinatoric_entropy}
\end{equation}
\end{widetext}
where the primed sums indicate constraint that $x_{A,B}$ are either both even or both odd. Fig.~\ref{fig:combinatoric_entropy} shows good agreement between Eq.~\eqref{eqn:combinatoric_entropy} and simulation-derived results, which also agree with the dynamics-derived entropy Eq.~(7) from the main text.

\section{Dilute-limit entropy scaling}
In the dilute-pair limit $p=1-\epsilon$, it is possible to utilize the known pair-length distribution to derive an asymptotically valid expression for entanglement entropy $S_A(N_A)$ in finite systems. Letting $N_A\leq N/2$ (which will always be true for at least one of the subsystems), we get
\begin{equation}
    S_A(N_A)
= \frac{\rho N}{Z_N} \bigg(
    \sum_{r=1}^{N_A} r^{1-\alpha}
    + N_A \sum_{r=N_A+1}^{N-N_A} r^{-\alpha}
    + \sum_{r=N-N_A+1}^{N-1} (N-r)r^{-\alpha}
\bigg),\label{eqn:SNA}
\end{equation}
in the case of open boundary conditions
where $\rho = \frac{1- n_1}{2}$ is the Bell pair density
and $Z_N=\sum_{r=1}^{N-1}(N-r)r^{-\alpha}$.
To derive large-$N$ entropy scaling from this equation, we utilize an Euler-Maclaurin expansion of the sum.
This is easily accomplished by expressing Eq.~\eqref{eqn:SNA} in terms of the expansion of harmonic numbers $H_n$ and generalized harmonic numbers $H_n^{(a)}$
\begin{align*}
H_n \equiv & H_n^{(1)}= \sum_{j=1}^n j^{-1} =
\log n + \gamma + \frac{1}{2n} + \mathcal O\left(n^{-2}\right),\\
H_n^{(a)}=&\sum_{j=1}^n j^{-a}=\zeta(a) + \frac{n^{1-a}}{1-a} + \frac{1}{2}n^{-a} + \mathcal O\left(n^{-a-1}\right),\quad a\neq 1,
\end{align*}
where $\zeta$ is the Riemann Zeta function and $\gamma$ is the Euler-Mascheroni constant $\gamma\approx 0.577$. To obtain the large-$N$ expansion, we then write \[
S_A(N_A)=\rho N
\frac{Z_N-Z_{N_A}-Z_{N-N_A}}{Z_N},
\]
where $Z_n=nH_n^{(\alpha)}-H_n^{(\alpha-1)}$. For $\alpha\neq1,2$, the Euler--Maclaurin expansion gives
\begin{equation}
    Z_n=n\zeta(\alpha)-\zeta(\alpha-1)
+\frac{n^{2-\alpha}}{(1-\alpha)(2-\alpha)}
+\mathcal O(n^{-\alpha}).
\end{equation}
 At $\alpha=1,2$, the corresponding expansions are
\begin{align*}
\left.Z_n\right|_{\alpha=1}
&=n(\log n+\gamma-1)+\frac12+\mathcal O(n^{-1}),\\
\left.Z_n\right|_{\alpha=2}
&=n\zeta(2)-\log n-\gamma-1+\mathcal O(n^{-2}),
\end{align*}

With $N_A=xN$ and assuming $N_A\leq N/2$ (which can always be made the case by switching which subsystem is labeled $A$), the leading-order contributions are
\begin{equation}
S_{A}^{(0)}(x) = \begin{cases} 
\rho f(x) N, & \alpha < 1 \\
\rho g(x) \frac{N}{\log N}, & \alpha = 1 \\
\frac{\rho f(x)}{(1-\alpha)(2-\alpha)\zeta(\alpha)} N^{2-\alpha}, &  1 < \alpha < 2 \\
\frac{\rho}{\zeta(2)}\log N, & \alpha = 2 \\
\frac{\rho \zeta(\alpha-1)}{\zeta(\alpha)}, & \alpha > 2.
\end{cases}
\label{eqn:entropy_leading}
\end{equation}
Spatial dependence is given by 
the functions
$f(x) = 1 - x^{2-\alpha} - (1-x)^{2-\alpha}$ and 
$g(x) = -x\log x - (1-x) \log(1-x)$.
The next terms in the expansion are
\begin{equation}
S_{A}^{(1)}(x) = \begin{cases} 
-\rho (1-\alpha)(2-\alpha)\zeta(\alpha)f(x) N^{\alpha}, & \alpha < 1 \\
\rho (1-\gamma) g(x) \frac{N}{\left(\log N\right)^2}, & \alpha = 1 \\
-\frac{\rho f(x)}{(1-\alpha)^2(2-\alpha)^2\zeta(\alpha)^2}N^{3-2\alpha}, &  1 < \alpha < 3/2 \\
\frac{\rho \zeta(\alpha-1)}{\zeta(\alpha)}-\frac{\rho f(x)}{(1-\alpha)^2(2-\alpha)^2\zeta(\alpha)^2}, & \alpha = 3/2 \\
\frac{\rho \zeta(\alpha-1)}{\zeta(\alpha)}, &  3/2 < \alpha < 2 \\
\frac{\rho}{\zeta(2)}\left[\log x(1-x) + 1 + \gamma \right], & \alpha = 2 \\
\frac{\rho f(x)}{(\alpha-1)(\alpha-2)\zeta(\alpha)}N^{2-\alpha}, & 2<\alpha <3,\\
\dfrac{\rho}{N}
\left[\dfrac{f(x)}{2\zeta(3)}
+\dfrac{\zeta(2)^2}{\zeta(3)^2}\right],& \alpha=3,\\
\dfrac{\rho}{N}
\left[\dfrac{\zeta(\alpha-1)}{\zeta(\alpha)}\right]^2, &\alpha>3.
\end{cases}
\label{eqn:entropy_subleading}
\end{equation}
Note the change in form at $\alpha=3/2$, which occurs when the exponent $N^{3-2\alpha}$ changes from positive to negative (with the overall scaling constant at $\alpha=3/2$).
For $\alpha>2$, the numerator's power-law correction is divided by the leading normalization $N\zeta(\alpha)$, while the constant term $-\zeta(\alpha-1)$ in $Z_N$ generates an additional positive $1/N$ correction. The $N^{2-\alpha}$ term dominates for $2<\alpha<3$, both terms contribute at $\alpha=3$, and the $1/N$ term dominates for $\alpha>3$.

\begin{figure}
\includegraphics[width=0.95\linewidth]{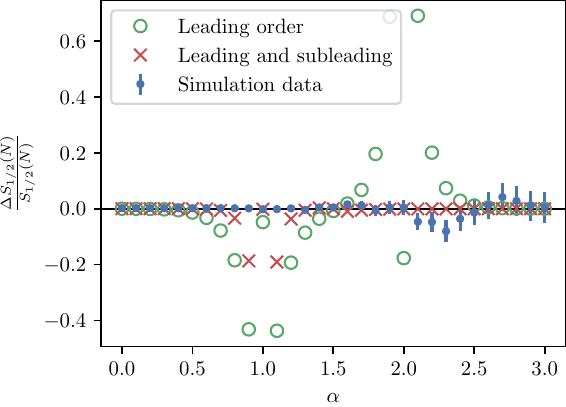}
\caption{Relative difference in entropy with respect to Eq.~\eqref{eqn:SNA} calculated at $p=0.95$ with $N=6400$.}
\label{fig:entropy_error}
\end{figure}

Fig.~\ref{fig:entropy_error} compares simulation data at $p=0.95$ to the entropies obtained from Eq.~\eqref{eqn:SNA} along with the approximations of Eq.~\eqref{eqn:entropy_leading} and Eq.~\eqref{eqn:entropy_subleading}. This comparison makes it clear that as $\alpha$ approaches the entanglement transitions at $\alpha=1$ and 2, higher-order terms in the expansion become increasingly relevant, which leads to significant deviations from clean power-law scaling of entropy.

\subsection{Periodic boundary conditions}
While we focus on open boundary conditions for the majority of our analysis, the same arguments used to derive Eq.~\eqref{eqn:SNA} may be applied with periodic boundary conditions. For the purposes of our pair probability distribution, we define distance as $r_{ij}=\min(|i-j|, N-|i-j|)$.
For $N$ even, this leads to pair distribution
\begin{equation}
P(r) = \frac{g(r) r^{-\alpha}}{Z_N}
\end{equation}
with $g(r) = 1$ for $r=N/2$ and $2$ otherwise. The normalization is
\begin{equation}
Z_N = \sum_{r=1}^{N/2} g(r) r^{-\alpha}
= 2 \sum_{r=1}^{N/2-1} r^{-\alpha} + \left(\frac{N}{2}\right)^{-\alpha}.
\end{equation}
Then, since the number of bonds of length $r$ connecting the subsystems is
\begin{equation}
C(N_A, r) = \begin{cases}
2\min(r, N_A, N-N_A) & 1 \leq r < N/2\\
\min(N_A, N-N_A) & r = N/2,
\end{cases}
\end{equation}
the average entanglement entropy is
\begin{equation}
S_A(N_A)=\frac{4\rho}{Z_N}\left(
\sum_{r=1}^{N_A} r^{1-\alpha}
+ N_A \sum_{r=N_A+1}^{N/2-1}r^{-\alpha}
+ \frac{N_A}{2} \left(\frac{N}{2}\right)^{-\alpha}
\right),
\end{equation}
for $N_A<N/2$, with the second sum dropped and the first sum truncated at $N/2-1$ in the case $N_A=N/2$.
The large-$N$ scaling of entropy from this equation takes the same form as the result of Eq.~\eqref{eqn:entropy_leading}.
\begin{figure}
\includegraphics[width=0.95\linewidth]{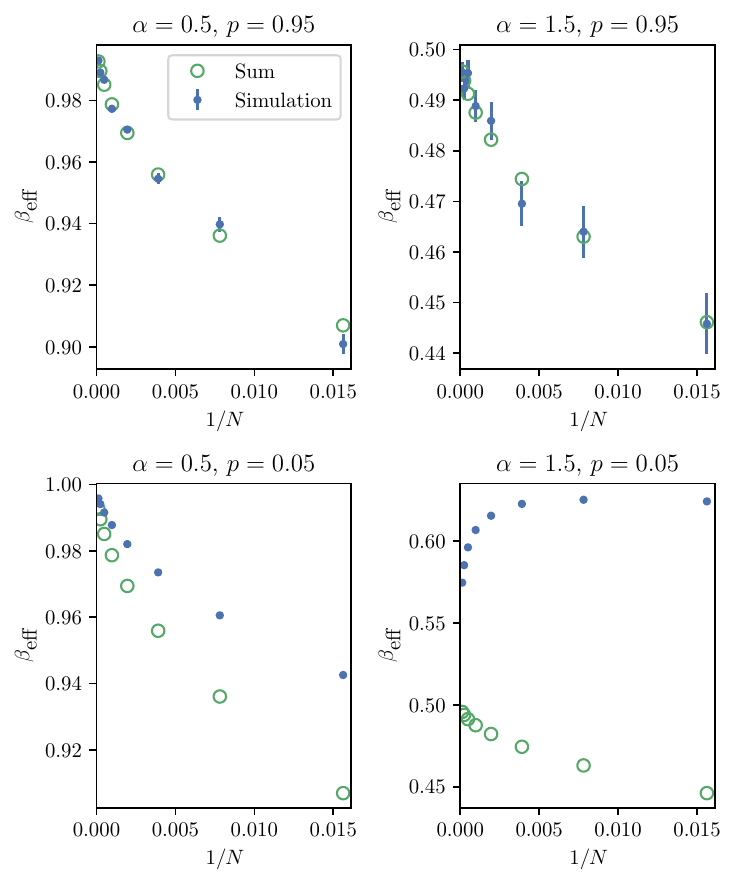}
\caption{Finite-size scaling of the effective exponent $\beta_\text{eff}$ defined in Eq.~\eqref{eqn:beta_eff}. Simulation results were obtained from $10^6$ trials on systems ranging from $N=32$ to $N=8192$. Error bars were obtained by propagating the SEM of $\braket{S_{A}(N/2)}$ simulation data. These errors are much smaller in the low-$p$ limit and are not visible on the plot. Circles correspond to values of $\beta_\text{eff}$ derived from Eq.~\eqref{eqn:SNA}.}
\label{fig:beta_eff}
\end{figure}

\section{Thermodynamic limit effective entropy exponents}

The large-$N$ expansion in Eqs.~\eqref{eqn:entropy_leading} and \eqref{eqn:entropy_subleading} indicates that significant finite-size corrections can obscure the asymptotic power-law entropy scaling
$S\sim N^\beta$, making it difficult to extract asymptotically-correct values of $\beta$ from finite systems.
To obtain thermodynamic-limit estimates, one approach is to perform finite-size scaling of the effective exponent 
\begin{equation}
    \beta_{\text{eff}}(N) = \frac{\log S_A(N/2) - \log S_{A}(N'/2)}{\log N - \log N'},
    \quad N' = \frac{N}{2}
    \label{eqn:beta_eff}
\end{equation}
which is a discretized approximation of $\beta(N) = \frac{d \log S_A(N/2)}{d\log N}$.
Fig.~\ref{fig:beta_eff} shows finite-size scaling of this quantity from simulations with $10^6$ trials compared to the result derived from Eq.~\eqref{eqn:SNA} at combinations of $\alpha = 0.5$, $1.5$ and $p=0.05$, $0.95$. The $p=0.95$ results show good agreement between the predicted and observed exponent as expected in the dilute-pair limit, with $\beta_\text{eff}\rightarrow 1$ for $\alpha=0.5$ and $\beta_{\text{eff}}\rightarrow 0.5=2-\alpha$ when $\alpha=1.5$.

At $p=0.05$, we no longer see good agreement with the $\beta_\text{eff}$ derived from Eq.~\eqref{eqn:SNA}, which is unsurprising since this equation should only be accurate in the dilute-pair limit.
Despite this, the $\alpha=0.5$ results remain consistent with volume-law scaling, with
$\beta_\text{eff}$ approaching 1 over the system sizes explored.

The behavior at $\alpha=1.5$ and $p=0.05$ is more interesting:
here, $\beta_\text{eff}$ decreases with increasing $N$. In the absence of a known finite-size scaling form for $\beta_\text{eff}$, we cannot reliably extract its $N\rightarrow\infty$ value. Nevertheless, the values over the system sizes studied remain in the range $0<\beta<1$, consistent with the fractal entropy phase.

\begin{figure}
\includegraphics[width=0.95\linewidth]{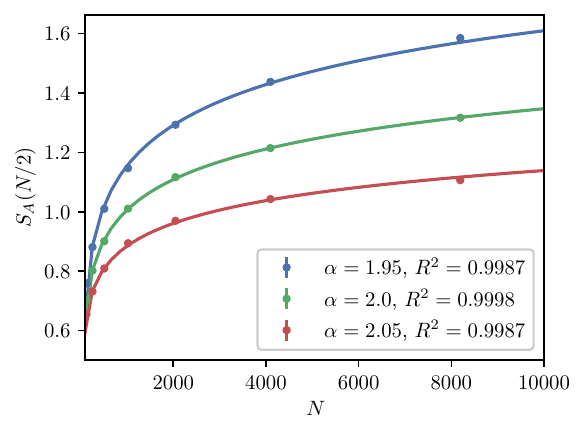}
\caption{Half-chain entropy obtained from simulation (dots) along with least-squares fits to a logarithmic function $A\log N + B$. This data was computed at $p=0.5$ and averaged over $10^5$ trials on system sizes ranging from $N=128$ to $8192$.}
\label{fig:alpha_c_example}
\end{figure}
\section{Entanglement transitions}
As discussed in the main text, finite-size corrections near the entanglement transitions at $\alpha_*$ and $\alpha_c$ cause significant deviations from clean power-law entropy scaling, making locating the transitions from fitted entropy exponent $\beta$ difficult. We instead identify the transitions using the characteristic finite-size scaling forms predicted at $\alpha=1,2$ in the dilute-pair limit (presented in Eqs.~\eqref{eqn:entropy_leading} and \eqref{eqn:entropy_subleading}). Fig.~\ref{fig:alpha_c_example} illustrates this procedure for $\alpha_c$ at $p=0.5$: fits of the half-chain entropy to $S_A(N/2)=A\log N + B$ maximize the coefficient of determination $R^2$ at $\alpha_c=2.00$. Repeating this procedure for each $p$ gives the transition locations shown in Fig.~2 of the main text.

\begin{figure}[t] 
\includegraphics[width=0.95\linewidth]{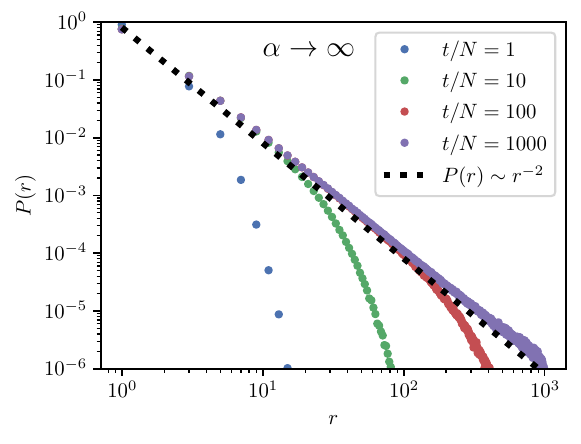}
\caption{Pair distribution resulting from simulations of an $N=1000$ qubit system with $p=0$ and in the limit $\alpha\rightarrow \infty$, where all Bell measurements are performed on nearest neighbor pairs. Since these dynamics produce exclusively odd-$r$ pairs, $P(r)$ is plotted only for odd $r$.
As time increases, the pair distribution becomes increasingly proportional to $r^{-2}$. This contrasts strongly with the input distribution (not shown), where $P(1)=1$ and $P(r>1)=0$.}
\label{fig:alpha_infty_distribution}
\end{figure}
\section{Dynamics and steady state at $p=0$}

\begin{figure}
\includegraphics[width=0.95\linewidth]{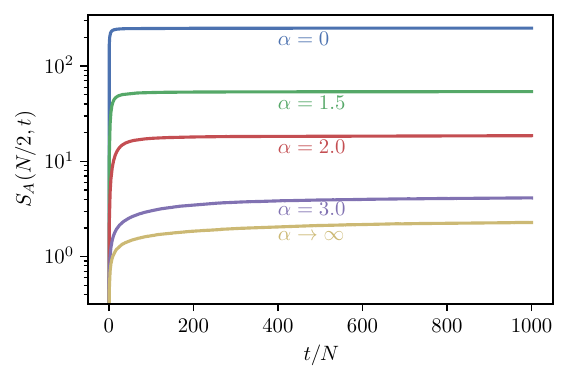}
\caption{Simulation-averaged half-chain entropy in a chain with $N=1000$ at $p=0$. Note the substantial change in relaxation dynamics as $\alpha$ increases.}
\label{fig:p=0_dynamics}
\end{figure}

\begin{figure}[!htbp]
\includegraphics[width=0.95\linewidth]{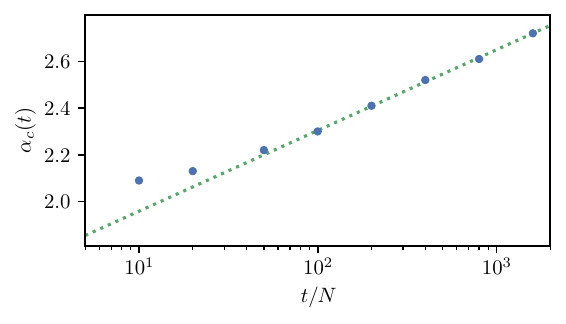}
\caption{Value of $\alpha$ at which the state best shows logarithmic entropy scaling as a function of time at $p=0$, showing approximately logarithmic time-dependence. The green line shows a least-squares fit of the form $A\log(t) + B$ performed using the data with $t/N\geq 100$.}
\label{fig:alpha_c_time_dep}
\end{figure}

The $p=0$ dynamics have well-understood limiting cases: at $\alpha=0$, our derived entropy remains volume-law. In the opposite limit $\alpha\rightarrow \infty$, our model shares the entanglement structure of the Nahum-Skinner Majorana loop model. In this model, entanglement swapping generates long pairs despite the measurements directly generating only nearest-neighbor pairs. In the steady state, the dynamics converge to a steady-state pair-length distribution $P(r)\sim r^{-2}$ and logarithmic entanglement scaling. Fig.~\ref{fig:alpha_infty_distribution} shows that our simulations approach this distribution at long times, providing a check that our model reproduces the known $\alpha\rightarrow\infty$ behavior.

Determining the behavior at finite $\alpha$ is complicated by slow relaxation dynamics. At $\alpha=p=0$, results presented in the main text show algebraic relaxation, while simulations at larger values of $\alpha$ indicate that relaxation slows further for $\alpha \gtrsim 1.5$ as shown in Fig.~\ref{fig:p=0_dynamics}.
For a finite observation time, applying the transition-locating procedure described in Sec.~V yields an apparent finite value $\alpha_c(t)$ separating fractal and area-law entanglement-scaling. However, unlike $\alpha_*$, which remains approximately stable over the accessible time window, the apparent boundary at $\alpha_c$ drifts towards larger $\alpha$ with increasing time. Fig.~\ref{fig:alpha_c_time_dep} shows the drift is approximately logarithmic over the range of times studied.

This drift of $\alpha_c(t)$, along with the known $\alpha\rightarrow\infty$ state, suggests the area-law phase vanishes for any finite $\alpha$ in the long-time limit. However, due to the slow relaxation dynamics, our simulations cannot distinguish whether $\alpha_c\rightarrow \infty$ or saturates at a finite value.